%% file: main.tex
\documentclass[aps,ap,twocolumn,showpacs,superscriptaddress,floatfix]{revtex4-2}
\usepackage{textgreek}
\usepackage{url}            
\usepackage{booktabs}       
\usepackage{amsfonts}       
\usepackage{nicefrac}       
\usepackage{microtype}      

\usepackage{graphicx}
\usepackage{natbib}
\usepackage{notoccite}

\usepackage{amsmath,amssymb,amsfonts}
\usepackage{graphicx}
\usepackage{acro}
\usepackage{siunitx}
\usepackage{subcaption}
\usepackage{float}
\usepackage{caption}
\usepackage{overpic}
\usepackage{multirow}
\usepackage{xspace}

\usepackage{wasysym}
\usepackage[hidelinks]{hyperref}
\usepackage[poorman,capitalise]{cleveref}
\usepackage{tikz}
\usepackage{pgfplots}
\usepackage{mathrsfs}
\usepackage{layouts}
\usepackage{placeins}

\input{tikzstuff}
\input{acro}

\input{newcmd}

\input{newunits}
\begin{document}
	\scapsetup
	\input{authors}
	\begin{abstract}
\begin{center}
			A wide-field magnetometer utilizing nitrogen-vacancy (NV) centers in diamond that does not require microwaves is demonstrated. It is designed for applications where microwaves need to be avoided, such as magnetic imaging of biological or conductive samples. The system exploits a magnetically sensitive feature of NV centers near the \ac{GSLAC}. An applied test field from a wire was mapped over an imaging area of \qty[parse-numbers=false]{\approx 500\times470}{\um\squared}. Analysis of the \ac{GSLAC} lineshape allows to extract vector information of the applied field.
The device allows micrometer-scale magnetic imaging at a spatial resolution dominated by the thickness of the NV layer (here $50\,\si{\um}$). For a pixel size of $4\,\si{\um} \times 3.8\,\si{\um}$ the estimated sensitivity is \qty{4.8}{\micro\tesla\per\sqrt\hertz}. Two modalities for visualizing the magnetic fields, static and temporal, are presented along with a discussion of technical limitations and future extensions of the method.
\end{center}
	\end{abstract}
	\maketitle

\section{Introduction}

Negatively charged \acf{NV} centers in diamond \cite{Maze2011} have attracted significant attention as a promising platform for sensing various physical quantities, such as temperature, pressure, magnetic and electric fields, at the nanoscale under various environmental conditions \cite{Lesik2019,Acosta2010,Jarmola2012}. Magnetometry based on spin-dependent fluorescence of these \ac{NV} centers has demonstrated sensitivities down to \qty[parse-numbers=false]{0.6}{\pico\tesla\per\sqrt\hertz} for ensembles at room temperature \cite{Chatzidrosos2017,Barry_2016,Barry2023}. By employing \acp{ODMR} with \ac{NV} centers, one can probe the magnetic fields generated by a wide range of samples, including biological systems, magnetic materials, current-carrying wires, and \acp{FPGA} \cite{Karadas2018,Barry_2016,Webb2021,Lenz2021a,Levine_2019}. In \ac{ODMR} based magnetometry, a diamond sample is illuminated with laser light (for example, at 532\,nm) and is subjected to a microwave field to drive population transfer between the differently bright spin states. When the microwave frequency matches a transition, there is a decrease of fluorescence. Measuring the resonant frequencies for transitions between the \mo and \mpm states allows reconstruction of the magnetic field projection onto the NV axis \cite{Jarmola2012,Doherty_2012,Ivady2021}.

However, there are scenarios where the use of microwaves is undesirable, such as when dealing with conductive materials or sensitive biological samples. To address this, recent developments have focused on microwave-free protocols that exploit energy-level crossings of \ac{NV} centers at different magnetic fields. These protocols include microwave-free magnetometry and vector magnetometry at zero-field, as well as the exploitation of specific features like the \ac{ESLAC} feature at 51.2\,mT and the \acf{GSLAC} feature at 102.45\,mT \cite{Wickenbrock_2016,Ivady2021,Auzinsh2018}. These kind of magnetically sensitive features have been successfully utilized for vector magnetometery, measuring eddy currents in conducting materials and performing \ac{NMR} \cite{Zheng2020,Zhang_2021,Wood2017}. In light of these advancements, this report presents a microwave-free wide-field magnetic microscope that leverages the \ac{GSLAC} feature to probe and image samples where microwaves are detrimental.

\begin{figure*}[htbp]
		\centering
        \includegraphics[width=1\linewidth]{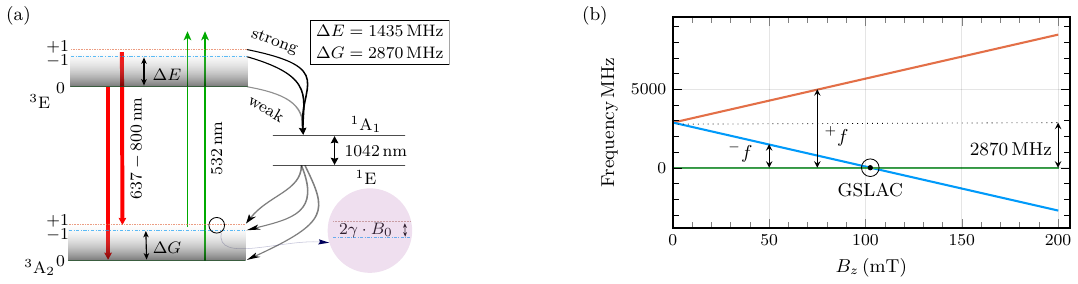}
		\caption{ (a) Energy levels of a NV$^-$ center \cite{Doherty_2012}
also showing the \ac{ISC} from $^3$E $\rightarrow$ $ ^1$A$_1$ $\rightarrow$ $^1$E $\rightarrow$ $^3$A$_2$ respectively. Two transition pathways, with and without the ISC, are indicated. The green and red arrows represent excitation and decay from and to the electronic ground state triplet; the black curves (``strong'' and ``weak'') represent decays via the \ac{ISC}. Strong and weak in this context means higher and lower rates of decay. The grey gradient is used to indicate the phonon sidebands of the respective triplets. (b) The eigenfrequencies of an NV center as a function of B$_z$ of the electron-spin magnetic sublevels of the ground state triplet neglecting hyperfine structure.}\label{fig:1}
	\end{figure*}

The \ac{GSLAC} feature originates from the mixing of ground-state levels at 102.45\,mT. Due to the Zeeman splitting of the triplet ground state, as illustrated in \cref{fig:1}a, the \msm state becomes degenerate with the \mo state and mixing results in population transfer between these states. Mixing occurs due to hyperfine interaction, cross-relaxation with other defects as well as static and oscillating transversal magnetic fields. The population transfer is observed as a drop in fluorescence intensity. Our study aims to explore the potential of utilizing \ac{NV} centers without the need for microwaves to map magnetic fields. We create static magnetic field maps by analyzing the lineshape of the \ac{GSLAC} resonance pixel by pixel. Furthermore, in another sensing modality, we investigate the sensitivity to dynamic field changes between individual image frames. 

To demonstrate the utility of this magnetometer we map the magnetic field distribution of a direct current (DC) field generated by a straight current-carrying wire with a diameter of \qty{200}{\um} placed on the diamond sample while we image it from underneath. This proof-of-concept experiment explores the application of \ac{NV} centers in a setup without microwaves. The lack of microwaves increases the usability of the device while reducing its technical complexity since no microwave components are needed. The results highlight the viability of diamond-based quantum sensors for a wide range of materials and biological applications.

\section{Experimental Setup}

	\begin{figure*}[htbp]
		\scapsetup
		\centering
		\includegraphics[width=\linewidth]{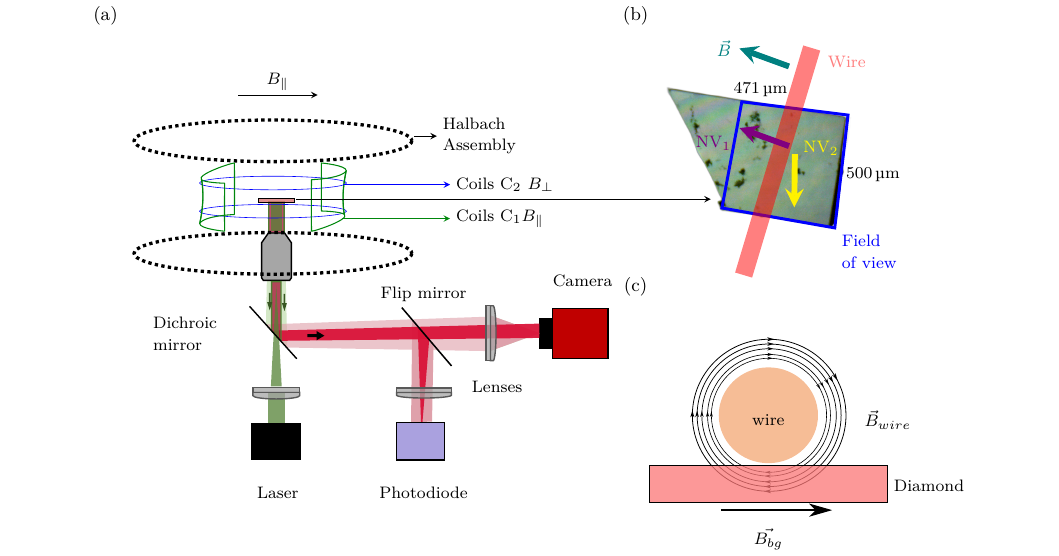}
		\caption{
			(a) The \ac{WF} imaging setup. It includes a diamond,an objective and the light path. Coils C$_1$ and C$_2$ are used to offset the axial and transverse magnetic fields. The dichroic mirror is used to separate the collected fluorescence from the reflected green light, and a flip mirror directs it either to a camera or a photodiode. The axis of the background and perpendicular magnetic field are denoted as \Bpar and \Bper respectively. (b) The [110] sample was made from a [100] cut diamond. The dark areas are presumably dirt within the crystal that is resistant to acid cleaning. The unreliable data from these positions were replaced with the mean of the five nearest reliable pixels. The thickness of the diamond is \qty{50}{\micro\meter} which sets the upper limit for spatial field resolution referred to in \cite{Scholten2021}. (c) An illustration of the magnetic field arrangement. The magnetic field by the wire $\vec{B}_\text{wire}$ counteracts $\vec{B}$ from the \acl{magnet} for positive currents.
		}\label{fig:2}
	\end{figure*}
	
	The setup incorporates a \acl{magnet} \cite{Wickenbrock_2021,Chatzidrosos_2021}, two pairs of coils, a (110) diamond plate and collection optics. The cylindrical \acl{magnet} provides a magnetic field orthogonal to its bore with minimal stray field external to the magnet. The diamond is a \ac{HPHT} grown diamond purchased from Element Six with a concentration of \qty{3.7}{\ppm} \ac{NV} centers homogeneously distributed. It was cut mechanically from a $^{13}$C depleted Ib diamond with a (100) face to a (110) face. The diamond sample dimensions were \qty[parse-numbers=false]{1 \times 0.5 \times 0.05}{\cubic\mm}. It had an initial nitrogen concentration $\le$ \qty{10}{\ppm}. It was irradiated with electrons of \qty{5}{\mega\eV}, with a dose of \qty{2e19}{\per\square\cm} electrons and then annealed at \qty{700}{\celsius} for \qty{8}{\hour}. The thickness of the NV layer (\qty{50}{\um}) limits the spatial magnetic field resolution \cite{Scholten2021}. The diamond is shown in \cref{fig:2}b where the field of view is marked in blue, the two NV axes of the diamond orthogonal to the cylinder bore usable for microwave-free magentometry are marked in yellow and violet. We use the orientation marked in violet.
	
	The sample is positioned in the center of the \acl{magnet} on a rotatable platform, with its axis of rotation parallel to the \acl{magnet} bore perpendicular to the field. This allowed to align the \ac{NV} axis in the (110) plane of the diamond to the \acl{magnet} background field produced. The sample is set on a sapphire window with a diameter of \qty{10}{\mm} and a thickness of 0.25\,mm.  It permits laser-light delivery and collection of light from under the diamond sample while acting as a heat sink.
	\begin{figure*}
		\centering
		\scapsetup
 		\includegraphics[width=1\linewidth]{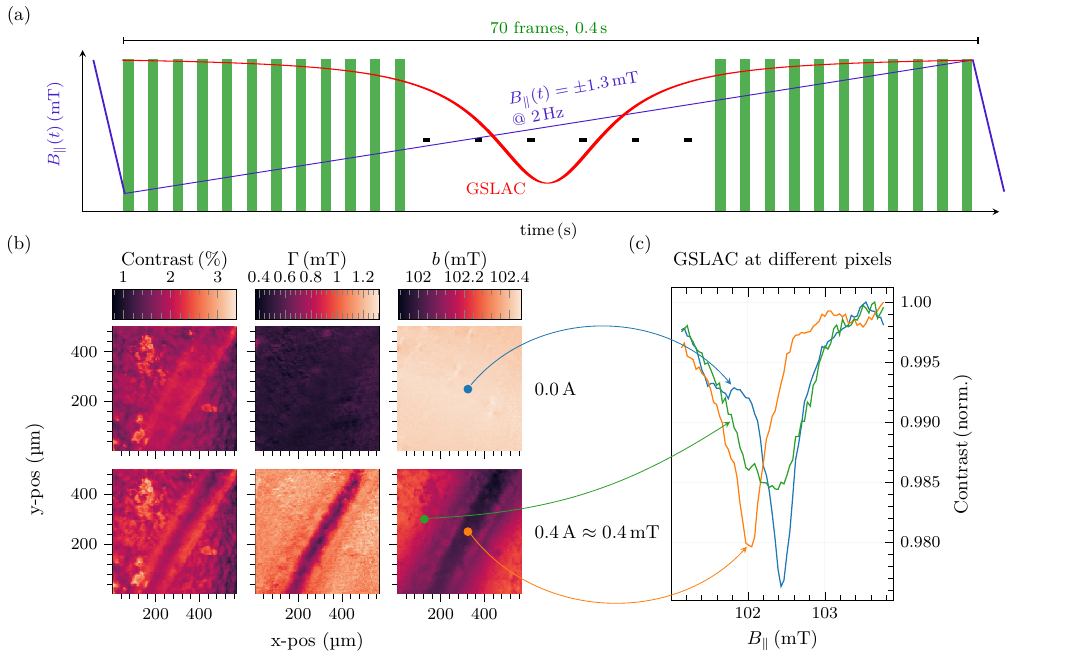}
		\caption{(a) Experimental sequence: data collection using a camera to capture pictures at the rising edges of 70 external triggers operating at \qty{185}{\hertz} are synchronized with a magnetic field sweep. The blue ramp indicated in (a) represents the field sweep, while the red trace illustrates the GSLAC feature. (b) Magnetic field maps obtained from a DC field generated with a wire carrying 0\,A and  0.4\,A, top and bottom, respectively. Each image is the result of the analysis of magnetic field scans in each pixel where the signal profiles are fit with a Lorentzian function. The contrast $C$, center field ($b$) and \ac{FWHM} ($\Gamma$) are then extracted and displayed. In (c) the \ac{GSLAC} was examined at two positions in the diamond and the results compared to the case without an applied test signal. The blue and orange traces are taken from the same position beneath the wire with and without a test signal. The green trace is for a position taken \qty{200}{\um} away from the wire center. This indicates that the axial component of the test field is more prominent closer to the wire and the transverse component is more prominent further away from the wire due to the curvature of the field consistent with the illustration in \cref{fig:2}c. This is also apparent \cref{fig:4}. 
}\label{fig:3}
	\end{figure*}
	As shown in \cref{fig:2} the \acl{magnet} is fitted with two sets of Helmholtz coil pairs. Coil pair C$_1$ is oriented in the direction of the background field and coil pair C$_2$ is perpendicular to it and parallel to the cylinder axis. The coils are used to shim the magnetic field of the \acl{magnet} and optimize the linewidth and contrast of the \ac{GSLAC} feature. The diamond mount is attached to the C$_1$ coils. It feature a range of \qty[parse-numbers=false]{\pm1.3}{\milli\tesla}. The second Helmholtz coil pair (C$_2$) is used to remove residual transverse fields. A third coil pair perpendicular to C$_1$ and C$_2$ would be ideal but was not implemented due to space constraints in the magnet bore.
	
	A \ac{cw} 532\,nm laser (Laser Quantum, gem532) is used to illuminate the diamond via a microscope objective (Olympus Plan 10X objective). The fluorescent light is gathered using the same objective, reflected off a short-pass dichroic mirror with a cutoff wavelength of 600\,nm  and passes through a long-pass filter to remove the remaining green reflection of the laser light. 

	The collected fluorescence is directed to a \acf{sCMOS} camera (Andor Zyla 5.5) for imaging. Alternatively for diagnostic purposes, the light can be directed to a photodiode (PDA36A2) using a flip mirror.
	\section{Results and Discussions}
	
	\subsection{Static Imaging}

	We demonstrate a static magnetic field imaging modality  using the \ac{GSLAC} by visualizing the field generated by a wire carrying electric current. The results are presented in \cref{fig:3}b
    To image , we illuminated the diamond with \ac{cw} 532\,nm laser light with a Gaussian profile. The laser beam covered an area of \qty[parse-numbers=false]{\approx 0.5\times0.47}{\mm\squared}.
    
	The background magnetic field was swept over a range of \qty[parse-numbers=false]{102.45\pm1.3}{\milli\tesla}, covering the \ac{GSLAC} feature while images where taken with the camera. This way the individual pixels contain a magnetic field scan over the \ac{GSLAC} feature. The current-carrying wire generates both axial (\Bpar) and transverse components (\Bper) in the measurement region, thereby leading to alterations in the contrast\,(C), \ac{FWHM} and the center-field position of the \ac{GSLAC} feature.
	
	To analyze and quantify the observed changes, we fitted the experimental data pixel-by-pixel with a Lorentzian function \cite{Petrakis_1967, Anishchik_2019} 
	to extract three fitting parameters: contrast ($C$), \ac{FWHM} ($\Gamma$), and center-field ($b$). 
	
	To ensure synchronization between the magnetic field ramp and the data acquisition, we employed an external trigger at 185\,Hz. This trigger served as a timing reference and triggered the camera to capture 70 frames during a 0.4\,s magnetic field ramp. 

	Additionally, to enhance the signal-to-noise ratio, we repeated the acquisition 20 times and averaged the results.The acquisition procedure, which is similar to that of \cite{Sengottuvel2022}, is illustrated in \cref{fig:3}a.
		
	During data analysis, we extract the \ac{GSLAC} feature from individual pixels within a specific region measuring \qty[parse-numbers=false]{\approx 0.47\times0.50}{\square\mm}. This region comprises a total of $625\times 715$\,pixels, with each pixel corresponding to an area of  $\approx 0.8\,\si{\um}\times0.69\,\si{\um}$ on the diamond surface. The extracted data are averaged over a $5\times5$\,pixels grid. This binning process reduces the optical resolution to approximately $\approx 4\,\si{\um} \times 3.4\,\si{\um}$ which is not limiting the spatial resolution for magnetic fields.
	
	We utilize the \ac{LM} algorithm \cite{Levenberg_1944} to fit the binned data on a pixel-by-pixel basis to a Lorentzian function. We extract the fitting parameters for each pixel, which provides information about the magnetic field distribution. The fitting of multiple pixels was done in parallel utilizing a threaded Python script. The resulting data-processing time was \qty{136}{\second} per image. This is still significantly longer than the data acquisition time and requires further improvement. For example, deploying a parallel fitting routine on a graphics card or \acp{FPGA}.
	
	The maps presented in \cref{fig:3}b are reconstructed images of a current-carrying wire. As the wire field in the sensor is mostly anti-aligned to the background field (see \cref{fig:2}c), the total magnetic field reduces. The wire field results in a shift and a broadening of the \ac{GSLAC} feature. On examination of the maps, we observe that the width of the feature is smallest at the center of the wire and increases away from it. This is mirrored in the center maps where the change in the total magnetic field is largest under the wire, \cref{fig:2}c and \cref{fig:3}c. This is consistent with a model of the wire magnetic field in the diamond.

In the region under the wire, there is no broadening of \ac{FWHM} caused by transverse fields. As we move away from the center of the wire, the transverse components become dominant, leading to an increase in the observed \ac{FWHM} of the \ac{GSLAC} feature. The center maps in \cref{fig:3}c also demonstrate this effect, with the area nearest to the wire showing a \qty{0.5}{\milli\tesla}-shift from \qty{102.4}{\milli\tesla} to \qty{101.9}{\milli\tesla} when a magnetic field corresponding to a current of 0.4\,A is applied. This shift gradually reduces to \qty{0.1}{\milli\tesla} furthest from the wire.

\begin{figure}[htbp]
		\centering
        \includegraphics[width=1\linewidth]{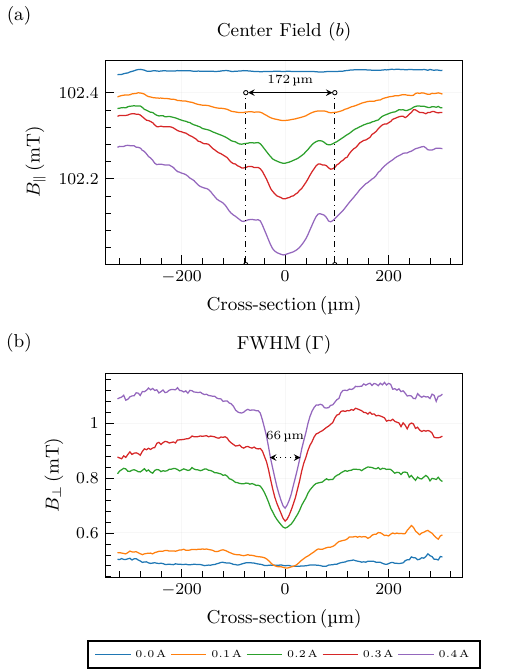}
		\caption{ 
			Cross-sections of the center and the \ac{FWHM} maps in \cref{fig:3}b displayed for multiple currents, respectively. The cross-section is perpendicular to the wire.  The center shift is maximal under the wire (a), while the width has a minimum there (b). This is expected from the pattern of the generated field as the width depends on the transverse component of the field illustrated in \cref{fig:2}c. 
			}\label{fig:4}
	\end{figure}

Additionally, we examine an averaged cross-section of the maps in \cref{fig:4}, specifically focusing on the \ac{FWHM} displayed in \cref{fig:4}b. We observe something peculiar: considering the thickness of the \ac{NV} layer in the diamond, which is approximately 50\,\si{\um}, the spatial resolution for the magnetic field was expected to be on the order of 50\,\si{\um} as well \cite{Scholten2021}. The wire itself has a diameter of 200\,\si{\um}. From observation however, the feature in the image \cref{fig:4}b  corresponding to the wire shows a \ac{FWHM} of \qty[parse-numbers=false]{\approx 66}{\um} when a field is generated with a wire carrying 0.4\,A. Just directly below the wire its field does not feature a transverse component in the NV layer. So just there, no broadening is observed. This effect becomes more pronounced with increasing wire current. It bears some resemblance to super-resolution imaging techniques like \ac{STED} \cite{Hell:94,Rittweger_2009}, albeit in the context of magnetic fields. In \ac{STED}, a laser beam with a donut-shaped cross-section is used to de-excite emitters through stimulated emission for sub-diffraction resolution images.

\subsection{Temporal Imaging}
	
\begin{figure*}[htbp]
 
		\includegraphics[width=1\linewidth]{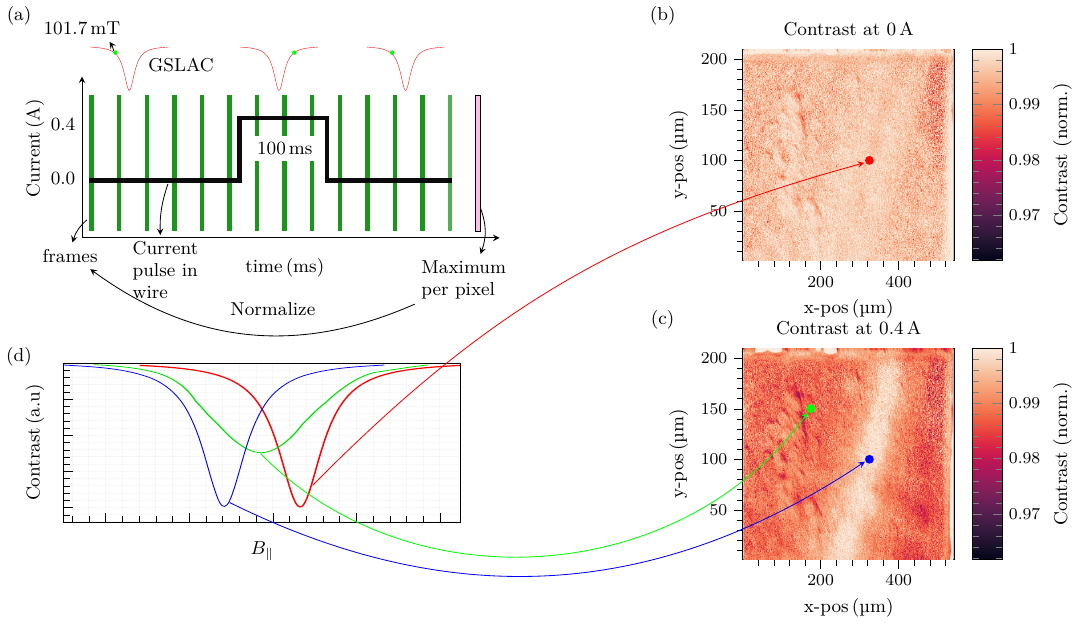}
		\caption{(a) Illustration of the data acquisition and processing sequence for temporal imaging. Fifteen frames are taken at 39\,Hz, while a current pulse is applied midst acquisition. The frames are normalized to the maximum value per pixel in order to visualize dynamics. In (b) and (c) we see the dynamics between two frames with a temporal separation of \qty{25}{\milli\second}. (b) The contrast with no test signal, while (c) displays the contrast variation induced when a current of \qty{0.4}{\ampere} generates a field. The field below the wire has been shifted to \qty{103.2}{\milli\tesla}, which causes it to appear brighter. The spatial resolution is restricted to \qty[parse-numbers=false]{\approx50}{\um}, which is the resolution limit for magnetic features that can be attained with the \samp. In the areas away from the wire the \ac{GSLAC} feature is broadened and shifted, which is observed as a reduction in contrast. (d) An illustration of the expected signal as a function of magnetic field at three different positions in the acquired frames for clarification of the involved processes.
		}\label{fig:5}
	\end{figure*}

To implement dynamic imaging, we captured a video while pulsing current through the wire. This allows us to observe the magnetic field from the wire in real time. The results are summarized in \cref{fig:5}. The temporal resolution is limited by the camera frame rate, which is set at 39\,Hz. The camera can operate at a significantly higher frame rate, however, this reduces the amount of collected photons.
	
	The experimental procedure is illustrated in \cref{fig:5}a. It involves tuning the bias field to the slope of the \ac{GSLAC} feature in  at \qty{101.7}{\milli\tesla}. We then capture 15 frames for \qty{380}{\milli\second} (at a rate of 39\,Hz) while turning on a current of 0.4\,A through the wire for the central 100\,ms. Post acquisition, we normalize the contrast of each individual pixel by its maximum value in the 15 images. This allows a straight-forward comparison of frames taken before and during the current pulse.
	
	In \cref{fig:5}b and \cref{fig:5}c), we present two frames separated by \qty{25}{\milli\second}, respectively. Figure \ref{fig:5}b has no current applied through the wire. In figure \ref{fig:5}c a current of 0.4\,A is applied. We observe changes in contrast in areas away from the wire. This is a result of the transverse components of the magnetic field (\Bper) generated by the wire. It broadens the \ac{GSLAC} feature. The broadening leads to a decrease in contrast and the corresponding diamond region to appear darker. However, directly under the wire, the magnetic field is aligned with the total field (\Bpar) causing the \ac{GSLAC} resonance to shift to \qty{103.2}{\milli\tesla} but not to broaden. This is why this region remains bright. We have chosen the values of the bias and the wire fields to maximize the contrast variation between the ``on'' and ``off'' frames integrated over the complete image. Furthermore, we chose these parameters to ensure all pixels remain within the \ac{GSLAC} feature and to determine the limit of contrast variations we can visualize.
	
\begin{figure*}[htbp]
		\scapsetup
        \includegraphics[width=1\linewidth]{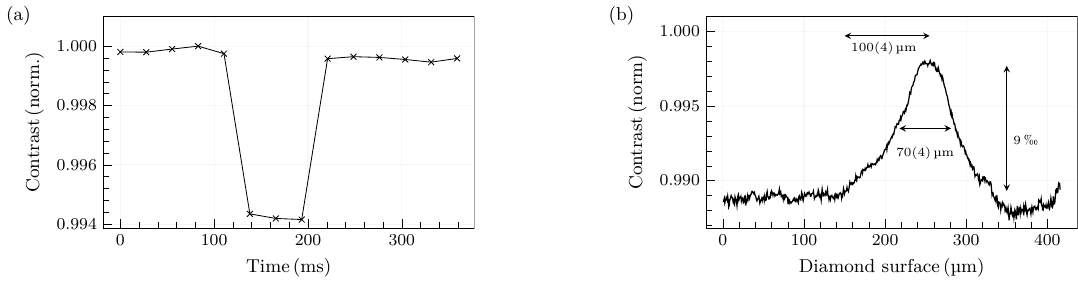}
		\caption{
			(a) Mean values of the averaged contrast per frame over the duration of acquisition, showing the applied pulse. (b) Mean values of the contrast for points in the direction orthogonal to the wire.
			We can detect a contrast drop of \qty{9}{\permille} up to \qty{100}{\um} from the center of the wire after which the contrast variation saturates.
		}\label{fig:6}
	\end{figure*}

	An analysis of the temporal cross-section of the acquired frames is shown in \cref{fig:6}a. The the applied pulse shape is retrieved. An examination of the cross-section of the wire image is presented in \cref{fig:6}b. We discern maximum contrast variations of up to \qty{9}{\permille} at a distance of \qty{100}{\um} from the wire center. Beyond this distance, the contrast variations, which is due mostly to the transverse field component of the wire, level off. To further improve the temporal and spatial resolution in this modality, high-frame-rate or lock-in detection-enabled cameras can be employed.

\section{Conclusion}

We developed a microwave-free diamond magnetic field microscope utilizing NV centers to measure spatially varying magnetic fields over a field of view of approximately \qty[parse-numbers=false]{500 \times 470}{\um}. We analyzed the GSLAC feature and can reconstruct longitudinal and transverse components of magnetic fields of samples. We implemented a fitting algorithm to extract the \ac{GSLAC} feature and estimate magnetic fields in an image of $125 \times 143$ pixels post-binning.

Our results demonstrate the potential of microwave-free vector magnetometry with temporal resolution enabling recording of magnetic ``movies''.
We envision that this approach will have a significant impact across a range of fields, including materials science, biology, and condensed matter physics.

It is important to acknowledge certain limitations in achieving high-resolution magnetic maps using the current methodology. One constraint is that only 1/4 of the NV-center orientations are available for this modality due to the requirement of on-axis alignment with the bias field. This leads to an overall drop in signal. Use of preferentially oriented NV centers could enhance the overall contrast and sensitivity \cite{Osterkamp_2019}.

Another limitation arises from the camera, which imposes constraints in terms of its Full Well Capacity (FWC) and frames per second (fps). The mean per pixel shot-noise limit is estimated to be approximately \qty{4.8}{\micro\tesla\per\sqrt\hertz} for a pixel size of \qty{4}{\um}, or \qty{137.35}{\micro\tesla\,{\raiseto{1.5}\um}\per\sqrt\hertz} for a volume-normalized estimate.
This can  be compared to the best-reported volume-normalized sensitivities of comparable cameras using microwaves (including lock-in cameras) of approximately \qty{31}{\nano\tesla\,{\raiseto{1.5}\um}\per\sqrt\hertz} \cite{Kazi_2021,Hart2021}. 
Considering these limits, future improvements  may involve lock-in-based detection to boost sensitivity and pulse sequences for T1 relaxometry \cite{Wood2017}.

\section{Acknowledgment}

This work was supported by the German Federal Ministry of Education and Research (BMBF) within the Quantumtechnologien program (DIAQNOS, project no. 13N16455) and by the European Commission’s Horizon Europe Framework Program under the Research and Innovation Action MUQUABIS, project no. 101070546.

	\bibliography{ref}
\bibliographystyle{ieeetr}
	
\end{document}

%% file: tikzstuff.tex
\usepgfplotslibrary{groupplots,dateplot}
\usetikzlibrary{patterns,shapes.arrows}
\pgfplotsset{compat=newest}
\usetikzlibrary{backgrounds}
\usepgfplotslibrary{patchplots}
\usepgfplotslibrary{fillbetween}
\pgfplotsset{%
	layers/standard/.define layer set={%
		background,axis background,axis grid,axis ticks,axis lines,axis tick labels,pre main,main,axis descriptions,axis foreground%
	}{grid style= {/pgfplots/on layer=axis grid},%
		tick style= {/pgfplots/on layer=axis ticks},%
		axis line style= {/pgfplots/on layer=axis lines},%
		label style= {/pgfplots/on layer=axis descriptions},%
		legend style= {/pgfplots/on layer=axis descriptions},%
		title style= {/pgfplots/on layer=axis descriptions},%
		colorbar style= {/pgfplots/on layer=axis descriptions},%
		ticklabel style= {/pgfplots/on layer=axis tick labels},%
		axis background@ style={/pgfplots/on layer=axis background},%
		3d box foreground style={/pgfplots/on layer=axis foreground},%
	},
	compat=newest,
	xticklabel style={font={{\fontsize{8 pt}{10.4 pt}\selectfont}},color={rgb,1:red,0.0;green,0.0;blue,0.0}, draw opacity={1.0}, rotate={0.0}},
	yticklabel style={font={{\fontsize{8 pt}{10.4 pt}\selectfont}},color={rgb,1:red,0.0;green,0.0;blue,0.0}, draw opacity={1.0}, rotate={0.0}},
	legend style={color={rgb,1:red,0.0;green,0.0;blue,0.0}, draw opacity={1.0}, line width={1}, solid, fill={rgb,1:red,1.0;green,1.0;blue,1.0}, fill opacity={1.0}, text opacity={1.0}, font={\fontsize{6 pt}{7.8 pt}\selectfont},
	},
	legend pos=outer north east,
	minor tick num=4,
	xmajorgrids={true},
	ymajorgrids={true},
	x grid style={color={rgb,1:red,0.0;green,0.0;blue,0.0}, draw opacity={0.1}, line width={0.5}, solid},
	y grid style={color={rgb,1:red,0.0;green,0.0;blue,0.0}, draw opacity={0.1}, line width={0.5}, solid},
	xtick align={inside},
	ytick align={inside},
	width=0.7\linewidth,
	height=0.7\linewidth,
	scale only axis,
	/pgf/number format/1000 sep={}
}


%% file: acro.tex
\DeclareAcronym{ODMR}{
	short = {ODMR},
	long = {optically detected magnetic resonance},
	tag = {abbrev}
}
\DeclareAcronym{NMR}{
	short = {NMR},
	long = {nuclear magnetic resonance},
	tag = {abbrev}
}
\DeclareAcronym{EM}{
	short = {EM},
	long = {electromagnetic radiation},
	tag = {abbrev}
}
\DeclareAcronym{GSLAC}{
	short = {GSLAC},
	long = {ground state level anticrossing},
	tag = {abbrev}
}
\DeclareAcronym{MRI}{
	short = {MRI},
	long =  {magnetic resonance imaging },
	tag = {abbrev}
}
\DeclareAcronym{PID}{
	short = {PID},
	long = {proportional-integral-derivative controller},
	tag = {abbrev}
}
\DeclareAcronym{FCC}{
	short = {FCC},
	long =  {face-centred cubic},
	tag = {abbrev}
}
\DeclareAcronym{NV}{
	short = {NV},
	long =  {nitrogen vacancy},
	tag = {abbrev}
}
\DeclareAcronym{-NV}{
	short = {NV$^-$},
	long =  {nitrogen vacancy},
	tag = {abbrev}
}
\DeclareAcronym{SG}{
	short = {SG},
	long = {signal generator},
	tag = {abbrev}
}
\DeclareAcronym{FG}{
	short = {FG},
	long = {function generator},
	tag = {abbrev}
}
\DeclareAcronym{MW}{
	short = {MW},
	long = {microwave},
	tag = {abbrev}
}
\DeclareAcronym{LCAO}{
	short = {LCAO},
	long = {linear combination of atomic orbitals} ,
	tag = {abbrev}
}
\DeclareAcronym{MO}{
	short = {MO},
	long = {molecular orbitals} ,
	tag = {abbrev}
}
\DeclareAcronym{ISC}{
	short = {ISC},
	long = {intersystem crossing },
	tag = {abbrev}
}
\DeclareAcronym{PL}{
	short = {PL},
	long = {photo-luminescence},
	tag = {abbrev}
}
\DeclareAcronym{cw}{
	short = {cw},
	long = {continuous wave},
	tag = {abbrev}
}
\DeclareAcronym{Mandhala}{
	short = {Mandhala},
	long = {magnetic arrangement for novel discrete halbach layout},
	tag = {abbrev}
}
\DeclareAcronym{HPHT}{
	short = {HPHT},
	long = {high-pressure high-temperature},
	tag = {abbrev}
}
\DeclareAcronym{SQ}{
	short = {SQ},
	long = {single quantum},
	tag = {abbrev}
}
\DeclareAcronym{DQ}{
	short = {DQ},
	long = {double quantum},
	tag = {abbrev}
}
\DeclareAcronym{fps}{
	short = {fps},
	long = {frames per second},
	tag = {abbrev}
}
\DeclareAcronym{magnet}{
	short = {magnet},
	long = {Halbach assembly},
	tag = {abbrev}
}
\DeclareAcronym{PM}{
	short = {PM},
	long = {cylindrical permanent magnet},
	tag = {abbrev}
}
\DeclareAcronym{LM}{
	short = {LM},
	long = {Levenberg-Marquardt},
	tag = {abbrev}
}
\DeclareAcronym{FWC}{
	short = {FWC},
	long = {full well capacity},
	tag = {abbrev}
}
\DeclareAcronym{FWHM}{
	short = {FWHM},
	long = {full width at half maximum},
	tag = {abbrev}
}
\DeclareAcronym{WF}{
	short = {WF},
	long = {wide-field},
	tag = {abbrev}
}
\DeclareAcronym{CVD}{
	short = {CVD},
	long = {chemical vapour deposition},
	tag = {abbrev}
}
\DeclareAcronym{MEG}{
	short = {MEG},
	long = {magnetoencephalography},
	tag = {abbrev}
}
\DeclareAcronym{MCG}{
	short = {MCG},
	long = {magnetocardiography},
	tag = {abbrev}
}
\DeclareAcronym{SQUID}{
	short = {SQUID},
	long = {superconducting quantum interference device},
	tag = {abbrev}
}
\DeclareAcronym{sCMOS}{
	short = {sCMOS},
	long = {scientific Complementary Metal-Oxide-Semiconductor},
	tag = {abbrev}
}
\DeclareAcronym{PSN}{
	short = {PSN},
	long = {photon shot-noise},
	tag = {abbrev}
}
\DeclareAcronym{OPM}{
	short = {OPM},
	long = {optically pumped magentometer},
	tag = {abbrev}
}
\DeclareAcronym{FPGA}{
	short = {FPGA},
	long = {field progammable gate array},
	tag = {abbrev}
}
\DeclareAcronym{ESLAC}{
	short = {ESLAC},
	long = {excited-state level anti-crossing},
	tag = {abbrev}
}
\DeclareAcronym{STED}{
	short = {STED},
	long = {stimulated emission depletion},
	tag = {abbrev}
}

%% file: newcmd.tex
\newlength{\totalimgwidth}
\newlength{\imgspacingwidth}

\DeclareMathSymbol{\shortminus}{\mathbin}{AMSa}{"39}
\newcommand{\mo}{$m_{s}=0$\xspace}
\newcommand{\mpm}{$m_{s}=\pm1$\xspace}

\newcommand{\msm}{$m_{s}=-1$\xspace}
\newcommand{\Bpar}{$B_{\|}$\xspace}
\newcommand{\Bper}{$B_{\bot}$\xspace}
\newcommand{\samp}{WF sample\xspace}

\newcommand{\scapsetup}{\captionsetup{subrefformat=parens}}



%% file: newunits.tex
\DeclareSIUnit{\gauss}{%
	\text{$\textup{G}$}%
}
\DeclareSIUnit{\ppm}{%
	\text{$\textup{ppm}$}%
}

\DeclareSIUnit\permille{\text{\textperthousand}}

%% file: authors.tex
\title{Microwave-free  wide-field magnetometry using nitrogen-vacancy centers}
\author{Joseph Shaji Rebeirro}
\affiliation{Johannes Gutenberg-Universit\"at Mainz, 55128 Mainz, Germany}
\affiliation{Helmholtz-Institut Mainz, GSI Helmholtzzentrum f{\"u}r Schwerionenforschung, 55128 Mainz, Germany}
\author{Muhib Omar}
\affiliation{Johannes Gutenberg-Universit\"at Mainz, 55128 Mainz, Germany}
\affiliation{Helmholtz-Institut Mainz, GSI Helmholtzzentrum f{\"u}r Schwerionenforschung, 55128 Mainz, Germany}
\author{Till Lenz}
\affiliation{Johannes Gutenberg-Universit\"at Mainz, 55128 Mainz, Germany}
\affiliation{Helmholtz-Institut Mainz, GSI Helmholtzzentrum f{\"u}r Schwerionenforschung, 55128 Mainz, Germany}
\author{Omkar Dhungel}
\affiliation{Johannes Gutenberg-Universit\"at Mainz, 55128 Mainz, Germany}
\affiliation{Helmholtz-Institut Mainz, GSI Helmholtzzentrum f{\"u}r Schwerionenforschung, 55128 Mainz, Germany}

\author{Peter Bl\"umler}
\affiliation{Johannes Gutenberg-Universit\"at Mainz, 55128 Mainz, Germany}

\author{Dmitry Budker}
\affiliation{Johannes Gutenberg-Universit\"at Mainz, 55128 Mainz, Germany}
\affiliation{Helmholtz-Institut Mainz, GSI Helmholtzzentrum f{\"u}r Schwerionenforschung, 55128 Mainz, Germany}
\affiliation{Department of Physics, University of California, Berkeley, CA 94720, USA}
\author{Arne Wickenbrock}
\email{wickenbr@uni-mainz.de}
\affiliation{Johannes Gutenberg-Universit\"at Mainz, 55128 Mainz, Germany}
\affiliation{Helmholtz-Institut Mainz, GSI Helmholtzzentrum f{\"u}r Schwerionenforschung, 55128 Mainz, Germany}

\date{\today}